\documentclass[preprint,preprintnumbers,amsmath,amssymb]{revtex4}
\usepackage{graphicx}
\usepackage{dcolumn}
\usepackage{bm}
\usepackage{epsfig}

\def\integral{\int_{-\infty}^{+\infty}}
\def\pre{\frac{1}{\sqrt{2\pi\hbar}}}
\begin{document}

\preprint{ {\bf February 2005}}

\title{Simple examples of position-momentum correlated 
Gaussian free-particle wavepackets in one-dimension 
with the general form of the time-dependent spread in position}

\author{R. W. Robinett}
\email{rick@phys.psu.edu}
\affiliation{
Department of Physics\\
The Pennsylvania State University\\
University Park, PA 16802 USA \\
}

\author{M. A. Doncheski} \email{mad10@psu.edu}
\affiliation{%
Department of Physics\\
The Pennsylvania State University \\
Mont Alto, PA 17237 USA \\
}

\author{L. C. Bassett} \email{lcb36@cam.ac.uk}
\affiliation{%
Department of Applied Mathematics and Theoretical Physics (DAMTP) \\
University of Cambridge \\
Wilberforce Road, Cambridge, CB3~0WA, United Kingdom \\
}

\date{\today}

\begin{abstract}
We provide simple examples of closed-form Gaussian wavepacket solutions 
of  the free-particle Schr\"{o}dinger equation in one dimension
which exhibit the most general form of the time-dependent spread in position,
namely 
$(\Delta x_t)^2 = 
(\Delta x_0)^2 + At + (\Delta p_0)^2t^2/m^2$,
where $A \equiv 
\left\langle 
(\hat{x}-\langle \hat{x} \rangle_0)
(\hat{p}-\langle \hat{p} \rangle_0) 
+ 
(\hat{p}-\langle \hat{p} \rangle_0)
(\hat{x}-\langle \hat{x} \rangle_0)
\right\rangle_0$ 
contains information on the position-momentum correlation
structure of the initial wave packet. We exhibit straightforward
examples corresponding to squeezed states, as well as quasi-classical cases, 
for which $A<0$ so that the position spread can (at least initially) decrease
in time because of such correlations. We discuss how the  initial 
correlations in these examples can be dynamically
generated (at least conceptually) in various bound state systems. 
Finally, we focus on providing different ways 
of visualizing the $x-p$ correlations present in these cases, including the 
time-dependent distribution of kinetic energy and the use of the Wigner 
quasi-probability distribution. We discuss similar results, both for the 
time-dependent $\Delta x_t$ and special correlated solutions, for the case 
of a particle subject to a uniform force.

\end{abstract}

\maketitle

\section{\label{sec_intro} Introduction}
\label{sec:introduction}

The study of time-dependent solutions of the one-dimensional
Schr\"{o}dinger equation is a frequent  topic in many
undergraduate textbooks on quantum mechanics. The problem of a Gaussian
or minimum-uncertainty wavepacket solution  for the case of a free particle
(defined more specifically below) is the most typical example cited, often 
being worked out in detail, or at least explored in problems \cite{texts}. 
The emphasis is often on the time-dependent position spread for such 
solutions, typically written in the forms
\begin{equation}
(\Delta x_t)^2 = 
(\Delta x_0)^2\left(1+\left(\frac{t}{t_0}\right)^2\right)
 =   (\Delta x_0)^2 + \frac{(\Delta p_0)^2 t^2}{m^2}
\label{not_general_case}
\end{equation}
where the spreading time or coherence time can be defined by $t_0 
\equiv m\Delta x_0/\Delta p_0$. Textbooks rightly point out the essentially
classical nature of much of this result, explained by the fact that 
the higher momentum components of the wave packet outpace the slower ones, 
giving a position-spread which eventually increases linearly with time as 
$\Delta x_t \approx \Delta v_0 t$,  where $\Delta v_0$ is identified with 
$\Delta p_0/m$.

The form of the expression for $\Delta x_t$ in Eqn.~(\ref{not_general_case})
is a special case of the most general possible form of the time-dependent
spatial width of a one-dimensional wave packet solution of the 
free-particle Schr\"{o}dinger equation which is well-known in the pedagogical
literature \cite{baird} - \cite{andrews}, but seemingly found in many fewer
textbooks \cite{merzbacher}. This general case can be written
in the form
\begin{equation}
(\Delta x_t)^2 = 
(\Delta x_0)^2 +
\left\langle 
(\hat{x}-\langle \hat{x} \rangle_0)
(\hat{p}-\langle \hat{p} \rangle_0) 
+ 
(\hat{p}-\langle \hat{p} \rangle_0)
(\hat{x}-\langle \hat{x} \rangle_0)
\right\rangle_0 \frac{t}{m}
+ \frac{(\Delta p_0^2) t^2}{m^2}
\label{general_case}
\end{equation}
where the coefficient of the term linear in $t$ measures a non-trivial 
correlation between the momentum- and position-dependence of the initial 
wave packet. 
While such correlations are initially not present in the standard Gaussian
wave packet example routinely used in textbook analyses, which therefore
gives rise to the simpler form in Eqn.~(\ref{not_general_case}), 
a non-vanishing $x-p$ correlation does develop for later times as has 
been discussed in at least
one well-known text \cite{bohm} and several pedagogical articles 
\cite{leblond}.

For wave packets which are constructed in such a way that large momentum 
components ($p > \langle \hat{p} \rangle_0$) are initially preferentially 
located in the `back' of the packet ($x < \langle \hat{x}\rangle_0$), 
the initial correlation can, in fact, be negative
leading to time-dependent wave packets which initially shrink in size,
while the long-time behavior of any 1D free particle wave packet is indeed 
always dominated
by the quadratic term in Eqn.~(\ref{general_case}), consistent with standard
semi-classical arguments. (We stress that we will consider here only 
localized wave packets which are square-integrable, for which the evaluation 
of $\Delta x_t$ and $\Delta p_t$ is possible, and not pure plane wave states 
nor the special non-spreading,  free-particle solutions discovered by Berry 
and Balazs \cite{berry}.)

For the standard Gaussian or minimum uncertainty wave packet used in most 
textbook examples, and in fact for any initial wave packet of the form 
$\psi(x,0) = R(x)\exp(ip_0(x-x_0)/\hbar)$ 
where $R(x)$ is a real function, this initial 
correlation vanishes and the more familiar special case of $\Delta x_t$
in Eqn.~(\ref{not_general_case}) results, leading many students to believe
that it  is the most general result possible. 
It is, however,  very straightforward to construct  initial quantum states 
consisting of simple Gaussian wave functions, such as squeezed states or 
linear combination of Gaussians, which have the required initial 
position-momentum correlations `built in', and which therefore exhibit 
the general form 
in  Eqn.~(\ref{general_case}), including examples where the position-space
wave packet can initially shrink in width. Since these examples can be 
analyzed with little or no more mathematical difficulty than the standard 
minimum-uncertainty cases commonly considered in textbooks \cite{texts}, 
we will focus on providing two such examples below. We will, however, also 
emphasize the utility of different ways of visualizing the time-dependent 
position-momentum correlations suggested by the form in 
Eqn.~(\ref{general_case}).

The derivation of Eqn.~(\ref{general_case}) has been most often
discussed \cite{baird}, \cite{styer} using the evaluation of the 
time-dependence of expectation values described by 
\begin{equation}
\frac{d}{dt} \langle \hat{A} \rangle
= \frac{i}{\hbar} \left\langle [\hat{H},\hat{A}] \right\rangle
\label{time-development}
\end{equation}
using the free particle Hamiltonian, $\hat{H} = \hat{p}^2/2m$,
or related matrix methods \cite{nicola}; since we are interested only
in expectation values of operators ($\hat{A} = \hat{x}$ or $\hat{p}$) 
which are themselves
independent of time, there is no additional $\langle d\hat{A}/dt\rangle$ term
in Eqn.~(\ref{time-development}).  In the next section, we
derive the necessary time-dependent expectation values of powers of
position and momentum 
in a complementary way, using very general momentum-space ideas.
(Identical methods can then also be used to evaluate the general form 
of $\Delta x_t$ for the related case of uniform acceleration, which we
discuss in Appendix~\ref{sec:appendix}.)
Then in Sec.~\ref{sec:standard} we briefly review the special case of the
minimum-uncertainty Gaussian wave packet (to establish notation) focusing
on the introduction of useful tools to help visualize possible 
correlations between position and momentum in free particle wave
packets, especially the direct visualization of the real/imaginary
parts of $\psi(x,t)$, the time-dependent spatial distribution of kinetic 
energy, as well as the Wigner quasi-probability distribution. 
Then, in Sec.~\ref{sec:correlated}, we exhibit two cases of 
correlated wave packets with the general form of $\Delta x_t$
in Eqn.~(\ref{general_case}), which are
simple extensions of these  standard results. A similar
example, involving squeezed states, has been discussed in 
Ref.~\cite{ford}, 
but we will focus here on understanding the detailed
position-momentum correlations which give rise to the term linear in 
$t$ in Eqn.~(\ref{general_case}), especially using the techniques
outlined in Sec.~\ref{sec:standard} for their visualization.
Finally, we make some concluding remarks as well as noting
in an Appendix that very similar results (both for the general form of 
the time-dependent $\Delta x_t$  and for the exemplary cases studied
here) can be obtained for the Schr\"{o}dinger equation corresponding to
the case of constant acceleration.

\section{Time-dependent $\Delta x_t$ using momentum-space wavefunctions}
\label{sec:momentum_space}

While the general result for the free-particle $\Delta x_t$ is most 
often obtained using formal methods involving the time-dependence of 
expectation values as in Eqn.~(\ref{time-development}), 
one can also evaluate time-dependent powers of position and momentum 
for a free particle in terms of the
initial wave packet quite generally in terms of the momentum-space
description of the quantum state, namely $\phi(p,t)$, obtaining the
same results, in a manner which is nicely complementary to more standard 
analyses. Depending on the ordering of topics in a given quantum mechanics 
course syllabus, this discussion  might well be applicable and understandable 
earlier in the curriculum than the more formal method. 

In this approach, the most general momentum-space wave function 
which solves the free-particle time-dependent Schr\"{o}dinger equation 
\begin{equation}
\frac{p^2}{2m}\phi(p,t) = \hat{H} \phi(p,t) =  \hat{E} \phi(p,t)
= i\hbar \frac{\partial}{\partial t} \phi(p,t)
\, , 
\end{equation}
can be written in the form 
\begin{equation}
\phi(p,t) = \phi_{0}(p)\, e^{-ip^2t/2m\hbar}
\end{equation}
with $\phi(p,0) = \phi_{0}(p)$ being the initial state wavefunction.
The $t$-dependent expectation values for powers of momentum are trivial 
since
\begin{eqnarray}
\langle \hat{p} \rangle_t & = &  \int_{-\infty}^{+\infty}
\, p \, |\phi_{0}(p)|^2\,dp \equiv  \langle \hat{p} \rangle_0 
\label{p_1} \\
\langle \hat{p}^2 \rangle_t & = &  \int_{-\infty}^{+\infty}
\, p^2 \, |\phi_{0}(p)|^2\,dp \equiv  \langle \hat{p}^2 \rangle_0 
\label{p_2}
\end{eqnarray}
so that 
\begin{equation}
(\Delta p_t)^2 = \langle \hat{p}^2\rangle_t - \langle \hat{p}\rangle_t^2
= 
\langle \hat{p}^2\rangle_0 - \langle \hat{p}\rangle_0^2
= 
(\Delta p_0)^2
\end{equation}
as expected for a free-particle solution for which 
$|\phi(p,t)|^2 = |\phi_{0}(p)|^2$ is independent of time. 

In this representation, the position operator is given by the 
non-trivial form $\hat{x} = i\hbar (\partial/\partial p)$, and 
the time-dependent 
expectation value of position can be written as
\begin{eqnarray}
\langle \hat{x} \rangle_t & = &
\int_{-\infty}^{+\infty}
[\phi(p,t)]^{*}\, \hat{x}\, [\phi(p,t)]\,dp \nonumber \\
& = & 
\int_{-\infty}^{+\infty}
\left[\phi_{0}^{*}(p)\,e^{+ip^2t/2m\hbar}\right]
\,
\left(i\hbar \frac{\partial}{\partial p}\right)
\left[\phi_{0}(p)\,e^{-ip^2t/2m\hbar}\right]\,
dp \nonumber \\
& = & 
\int_{-\infty}^{+\infty}
[\phi_{0}^{*}(p)] \left(i\hbar \frac{\partial }{\partial p}\right)
[\phi_{0}(p)]\,dp
+ \frac{t}{m} \int_{-\infty}^{+\infty} \, p\, |\phi_{0}(p)|^2\,dp
\nonumber \\
& = & \langle \hat{x} \rangle_0 + \frac{t}{m} \langle \hat{p}\rangle_0
\label{x_1}
\end{eqnarray}
which is consistent with Ehrenfest's theorem for the essentially
classical behavior of $\langle \hat{x}\rangle_t$. 
The same formalism can
be used to evaluate $\langle \hat{x}^2\rangle_t$ and gives
\begin{equation}
\langle \hat{x}^2\rangle_t
 = 
\langle \hat{x}^2 \rangle_0
+ \frac{t}{m} \langle \hat{x} \hat{p} + \hat{p} \hat{x}\rangle_0
+ \langle \hat{p}^2\rangle_0 \frac{t^2}{m^2}
\label{x_2}
\end{equation}
where one can use the general representation-independent commutation
relation $[\hat{x},\hat{p}] = i\hbar$ to simplify the answer to this form.
The symmetric combination of position
and momentum operators, written here as $(\hat{x}\hat{p}+\hat{p}\hat{x})$, 
which is obviously Hermitian, guarantees that this expression is manifestly 
real.
(Discussions in textbooks on symmetrizing products of non-commuting operators
abound, but such results are seldom put into the context of being useful
or natural in specific calculations, as is apparent in their use here.)

Combining Eqns.~(\ref{x_1}) and (\ref{x_2}) then gives the most general
form for the time-dependent spread in position to be
\begin{eqnarray}
(\Delta x_t)^2 & = & \langle \hat{x}^2\rangle_t 
- \langle \hat{x}\rangle_t^2 \nonumber \\
& = & 
\left(\langle \hat{x}^2 \rangle_0
+ \frac{t}{m} \langle \hat{x} \hat{p} + \hat{p} \hat{x}\rangle_0
+ \langle \hat{p}^2\rangle_0 \frac{t^2}{m^2} \right)
\nonumber 
- \left(\langle \hat{x} \rangle_0 + \frac{t}{m} \langle \hat{p}\rangle_0\right)^2
\nonumber \\
& = & 
(\Delta x_0)^2 +
\left(
\langle \hat{x} \hat{p} + \hat{p} \hat{x} \rangle_0
- 2 
\langle \hat{x} \rangle_0 \langle \hat{p} \rangle_0
\right)
\frac{t}{m}
+ \frac{(\Delta p_0^2) t^2}{m^2} 
\nonumber \\
& = & 
(\Delta x_0)^2 +
\left\langle 
(\hat{x} - \langle \hat{x} \rangle_0)
(\hat{p} - \langle \hat{p} \rangle_0) 
+ 
(\hat{p} - \langle \hat{p} \rangle_0)
(\hat{x} - \langle \hat{x} \rangle_0)
\right\rangle_0 
\frac{t}{m}
+ \frac{(\Delta p_0^2) t^2}{m^2} 
\, . 
\end{eqnarray}
We have rewritten 
the term linear in $t$ in a form which stresses that it is a correlation 
between $x$ and $p$, similar in form to related classical quantities 
such as the covariance in probability and statistics. Recall that for
two classical quantities, $A$ and $B$, described by a joint probability 
distribution, the covariance is defined as
\begin{equation}
cov(A,B) = 
\left \langle
\left( A - \langle A \rangle \right)
\left( B - \langle B \rangle \right)
\right \rangle
= \langle AB \rangle - \langle A \rangle \langle B \rangle
\,. 
\end{equation}
As we will see in the next section, there is no initial correlation for 
the familiar minimum-uncertainty Gaussian wave packets. However, for simple 
variations on the standard example, as in Sec.~\ref{sec:correlated}, we will
find non-vanishing correlations, which we can visualize with the methods in
Sec.~\ref{sec:standard}.

We stress  that the notion of a time-dependent correlation between $x$ and 
$p$ at arbitrary times ($t>0)$ can be easily generalized from these results, 
and we can define a generalized covariance for these two variables 
\cite{merzbacher} -- \cite{leblond} 
(or any two operators, $\hat{A}, \hat{B}$) as
\begin{equation}
cov(\hat{x},\hat{p};t) \equiv \frac{1}{2} 
\left\langle 
(\hat{x} - \langle \hat{x} \rangle_t)
(\hat{p} - \langle \hat{p} \rangle_t) 
+ 
(\hat{p} - \langle \hat{p} \rangle_t)
(\hat{x} - \langle \hat{x} \rangle_t)
\right\rangle_t
\label{covariance}
\end{equation}
where the additional factor of $1/2$ accounts for the necessarily
symmetric combination which appears, compared to the classical
definition. One can then speak
of a time-dependent correlation coefficient defined by
\begin{equation}
\rho(x,p;t) \equiv
\frac{cov(x,p;t)}{\Delta x_t\cdot \Delta p_t}
\label{correlation_coefficient}
\end{equation}
in analogy with related quantities from statistics. This correlation
can be shown \cite{leblond} to satisfy the inequality
\begin{equation}
[\rho(x,p;t)]^2 \leq 1 - \left(\frac{|\langle [\hat{x},\hat{p}]\rangle|}{2\Delta x_t
\cdot \Delta p_t}\right)^2
= 1- \left(\frac{\hbar}{2\Delta x_t\cdot \Delta p_t}\right)^2
\end{equation}
which vanishes for the standard minimum-uncertainty Gaussian
at $t=0$, but which is non-zero for later times,  as we will see below.

\section{Standard minimum-uncertainty Gaussian wave packets}
\label{sec:standard}

The standard initial minimum-uncertainty Gaussian wave packet, which gives 
the familiar time-dependent spread in Eqn.~(\ref{not_general_case}), 
can be  written in generality as 
\begin{equation}
\phi_0(p) = \phi_{(G)}(p,0) = 
\sqrt{\frac{\alpha}{\sqrt{\pi}}}
\; e^{-\alpha^2(p-p_0)^2/2}
\; e^{-ipx_0/\hbar}
\label{initial_gaussian}
\end{equation}
where $x_0,p_0$ are used to characterize the arbitrary initial central 
position and momentum values respectively. This form gives 
\begin{equation}
\langle \hat{p} \rangle_{t} = p_0
\, , 
\qquad
\quad
\langle \hat{p}^2 \rangle_{t} = p_0^2 + \frac{1}{2\alpha^2}
\, ,
\qquad
\mbox{and}
\qquad
\Delta p_t = \Delta p_0 = \frac{1}{\alpha \sqrt{2}}
\label{momentum_results}
\end{equation}
which are, of course, consistent with the general results in 
Eqns.~(\ref{p_1}) and (\ref{p_2}).

The explicit form of the position-space wave function is given 
by Fourier transform as 
\begin{equation}
\psi_{(G)}(x,t) = \pre \sqrt{\frac{\alpha}{\sqrt{\pi}}}
\integral\, e^{ip(x-x_0)/\hbar}\,
e^{-\alpha^2 (p-p_0)^2/2}\,
e^{-ip^2t/2m\hbar}\,dp
\end{equation}
which can be evaluated in closed form (using the change of variables
$q \equiv p-p_0$ and standard integrals) to obtain
\begin{equation}
\psi_{(G)}(x,t) = \frac{1}{\sqrt{\sqrt{\pi} \alpha \hbar (1+it/t_0)}}
\,
e^{ip_0(x-x_0)/\hbar}
\, e^{-ip_0^2t/2m\hbar}
\,
e^{-(x-x_0-p_{0}t/m)^2/2(\alpha \hbar)^2(1+it/t_0)}
\label{free_particle_position_solution}
\end{equation}
where $t_0 \equiv m\hbar \alpha^2$ is the spreading time.  
This then gives 
\begin{equation}
|\psi_{(G)}(x,t)|^2 = \frac{1}{\sqrt{\pi}\beta_t}
\, e^{- [x-\overline{x}(t)]^2/\beta_t^2}
\end{equation}
where 
\begin{equation}
\overline{x}(t) \equiv x_0 + p_0t/m
\qquad
\mbox{and}
\qquad
\beta_t \equiv  \beta  \sqrt{1+(t/t_0)^2}
\qquad
\mbox{with}
\qquad
\beta \equiv \alpha \hbar
\,. 
\end{equation}
This gives
\begin{equation}
\langle \hat{x} \rangle_t = \overline{x}(t)
\quad
\qquad
\mbox{and}
\qquad
\quad
\langle \hat{x}^2 \rangle_t = [\overline{x}(t)]^2 + \frac{\beta_t^2}{2},
\end{equation}
so that
\begin{equation}
(\Delta x_t)^2 = 
\frac{\beta_t^2}{2}
= 
\frac{\beta^2}{2}
\left(1+\left(\frac{t}{t_0}\right)^2\right)
=  (\Delta x_0)^2 + (\Delta p_0 t/m)^2
\label{gaussian_result}
\end{equation}
which is the familiar textbook result, and for $t=0$ has the minimum
uncertainty product $\Delta x_0 \cdot \Delta p_0 = \hbar/2$. 

It is easy to confirm by direct calculation that there is no initial 
($t=0$) $x-p$ correlation ($cov(x,p;0)=0$) for this wavefunction, 
consistent with
the lack of a term linear in $t$ in Eqn.~(\ref{gaussian_result}). We 
emphasize that such correlations do indeed develop as the wavepacket 
evolves in time, which can be seen by examining the form of either the real or
imaginary parts of $\psi_{(G)}(x,t)$ as shown in Fig.~1 (where we specify
the model parameters used in that plot in the accompanying figure caption). 
We note that for times $t> 0$, the `front end' of the wave packet shown 
there is clearly more `wiggly' than the `back end' (simply count the nodes
on either side of $\langle x \rangle_t$.)
The time-dependent correlation function or covariance defined 
in Eqn.~(\ref{covariance}) and correlation coefficient
from Eqn.~(\ref{correlation_coefficient}) are easily calculated 
for this specific case to be
\begin{equation}
cov(x,p;t) = \frac{\hbar}{2} \left(\frac{t}{t_0}\right)
\qquad
\quad
\mbox{and}
\quad
\qquad
\rho(x,p;t) = \frac{t/t_0}{\sqrt{1+(t/t_0)^2}}
\label{standard_gaussian_correlations}
\end{equation}
which clearly expresses the increasingly positive correlation of
fast (slow) momentum components being preferentially in the leading 
(trailing) edge of the wave packet. We note that such correlations
have been discussed in Refs.~\cite{bohm} and \cite{leblond}.

This observation can also be described quantitatively by examining the 
distribution of kinetic energy of such a free-particle Gaussian wavepacket 
\cite{bassett}. 
In this approach, the standard expression for the kinetic energy is 
rewritten using  integration-by-parts in the form
\begin{equation}
\langle \hat{T}\rangle_{t}
 =  \frac{1}{2m}\langle \hat{p}^2\rangle_{t}
 =  -\frac{\hbar^2}{2m}
\integral dx \,\psi^*(x,t) \frac{\partial ^2 \psi(x,t)}{\partial x^2} 
 =  \frac{\hbar^2}{2m}\integral dx 
\left|\frac{\partial \psi(x,t)}{\partial x}\right|^2 
\end{equation} 
which can be used to define a {\it local kinetic energy density}, 
${\cal T}(x,t)$, via
\begin{equation}
{\cal T}(x,t) \equiv 
\frac{\hbar^2}{2m} 
\left|\frac{\partial \psi(x,t)}{\partial x} \right|^2
\qquad
\quad
\mbox{where}
\qquad
\quad
\langle \hat{T} \rangle_t =  \integral {\cal T}(x,t)\,dx 
\equiv T(t)
\, .
\label{kinetic_energy_distribution}
\end{equation}
As this notion is useful in systems other than for free particle states,
we allow for the possibility that the total kinetic energy varies with
time. 
Since this local density is clearly real and positive-definite, we can use it
to visualize the distribution of kinetic energy (or `wiggliness') 
in any time-dependent wavefunction. We can then define similar quantities
for the kinetic energy in the `front' and/or `back' halves of the wave
packet, using $\langle x\rangle_t$ as the measuring point, via
\begin{equation}
T^{(+)}(t) \equiv  \int_{\langle x \rangle_t}^{+\infty} {\cal T}(x,t)\,dx 
\qquad
\quad
\mbox{and}
\quad
\qquad
T^{(-)}(t) \equiv  \int^{\langle x \rangle_t}_{-\infty} {\cal T}(x,t)\,dx 
\label{half_kinetic_energies}
\, . 
\end{equation}

For the standard Gaussian wave packet in 
Eqn.~(\ref{free_particle_position_solution}), the local kinetic energy 
density is given by
\begin{equation}
{\cal T}_{(G)}(x,t) = \frac{1}{2m}
\left( p_0^2 + \left[\frac{2[x-\overline{x}(t)] p_0}{\alpha^2\hbar}\right]
\left[\frac{t/t_0}{(1+t^2/t_0^2)}\right]
+ \frac{[x-\overline{x}(t)]^2}{(\alpha^2 \hbar)^2 (1+t^2/t_0^2)}\right)
|\psi_{(G)}(x,t)|^2
\, . 
\label{gaussian_case}
\end{equation}
The expectation value of the kinetic energy is correctly given by
\begin{equation}
T_{(G)}(t) = \integral\, {\cal T}_{(G)}(x,t)\,dx = \frac{1}{2m} 
\left(p_0^2 + \frac{1}{2\alpha^2}\right)
\end{equation}
and receives non-zero contributions from only the first and last terms in 
brackets
in Eqn.~(\ref{gaussian_case}),  since the term linear in 
$[x-\overline{x}(t)]$ 
vanishes (when integrated over all space) for symmetry reasons.  The
individual values of $T^{(\pm)}_{(G)}(t)$ can also be calculated and 
are given by
\begin{equation}
T^{(\pm)}_{(G)}(t) 
= \frac{1}{2m}
\left(\frac{1}{2}\right)
\left( 
p_0^2 
\pm 
\left(\frac{2p_0}{\alpha \sqrt{\pi}} \right) \frac{t/t_0}{\sqrt{1+t^2/t_0^2}} 
+ \frac{1}{2\alpha^2}
\right)
\label{left_and_right_kinetic_energies}
\end{equation}
which are individually positive definite. The time-dependent fractions
of the total kinetic energy contained in the $(+)/(-)$ (right/left) halves 
of this standard wave packet are given by
\begin{equation}
R^{(\pm)}_{(G)}(t) \equiv 
\frac{T^{(\pm)}_{(G)}(t)}{T^{(+)}_{(G)}(t) + T^{(-)}_{(G)}(t)}
= \frac{1}{2} \pm 
 \left(\frac{2}{\sqrt{\pi}}\right)
\left( \frac{(p_0\alpha)}{(2(p_0\alpha)^2+1)}\right) 
\frac{t/t_0}{\sqrt{1+t^2/t_0^2}}
\label{define_r_function}
\,. 
\end{equation}
For the model parameters used in Fig.~1, for $t=2t_0$ this corresponds 
to $R^{(+)}/R^{(-)} = 56\%/44\%$, consistent with the small, but obvious,
difference in the kinetic energy distribution seen by `node counting'.

Finally, this growing correlation can be exhibited in yet another way, 
namely through the Wigner quasi-probability distribution, defined by
\begin{eqnarray}
P_{W}(x,p;t)
 & \equiv &
\frac{1}{\pi \hbar}
\int_{-\infty}^{+\infty}
\psi^{*}(x+y,t)\,\psi(x-y,t)\,e^{+2ipy/\hbar}\,dy \\
& = & 
\frac{1}{\pi \hbar}
\int_{-\infty}^{+\infty}
\phi^*(p+q,t)\, \phi(p-q,t)\, e^{-2ixq/\hbar}\,dq
\label{wigner_function}
\, . 
\end{eqnarray}
This distribution, first discussed by Wigner \cite{wigner}, 
and reviewed extensively in the research \cite{wigner_research}
and pedagogical \cite{wigner_pedagogical} literature (and even in the 
context of wave packet spreading \cite{wigner_lee}),
is as close as one can come to a quantum phase-space distribution,
and while not directly measurable, can still be profitably used to 
illustrate any $x-p$ correlations. 
For the standard minimum-uncertainty Gaussian wavefunctions defined by 
Eqns.~(\ref{initial_gaussian}) or (\ref{free_particle_position_solution}), 
one finds that \cite{kim_noz}
\begin{equation}
P_{W}(x,p;t) = \frac{1}{\hbar \pi}
\, e^{-(p-p_0)^2 \alpha^2}
\, e^{-(x-x_0-pt/m)^2/\beta^2}
= 
P_{W}(x-pt/m,p;0)
\,.
\label{explicit_wigner_function}
\end{equation}
Contour plots of $P_{W}(x,p;t)$ corresponding to the time-dependent
standard Gaussian wave packet for two different times ($t=0$ and
$t=2t_0$) are also shown at the bottom of Fig.~1, where the the 
elliptical contours with principal axes parallel to the $x,p$ 
axes for the $t=0$ case are indicative of the vanishing initial correlation, 
while the slanted contours at later times are consistent with the correlations
developing as described by Eqn.~(\ref{standard_gaussian_correlations}).
(We note that Bohm \cite{bohm} uses a similar illustration, but discusses it 
only in the context of classical phase space theory and Liouville's theorem.)
The visualization tools used in Fig.~1 (explicit plots of 
$Re[\psi(x,t)]$, and the Wigner function) and the distribution of
kinetic energy as encoded in Eqns.~(\ref{left_and_right_kinetic_energies})
or (\ref{define_r_function}), 
can then directly be used to examine the correlated wave packets we 
discuss in the next section.

As a final reminder about the quantum mechanical ``engineering''
of model one-dimensional wavepackets, we recall that since an initial
$\phi_{0}(p)$ is related to the time-dependent $\psi(x,t)$ for
free-particle solutions via
\begin{equation}
\psi(x,t) = \frac{1}{\sqrt{2\pi\hbar}}\,
\int_{-\infty}^{+\infty}\,
\left[\phi_{0}(p)\,e^{-ip^2t/2m\hbar}\right]\,e^{ipx/\hbar}\,dp
\end{equation}
then the simple modification 
\begin{equation}
\tilde{\phi}_{0}(p) = \phi_{0}(p)\, e^{-ipa/\hbar}\,e^{ip^2\tau/2m\hbar}
\label{change_phi}
\end{equation}
leads to the related position-space wavefunction satisfying
\begin{equation}
\tilde{\psi}(x,t)  = \frac{1}{\sqrt{2\pi\hbar}}\,
\int_{-\infty}^{+\infty}\,
\left[\left(\phi_{0}(p)\, e^{-ipa/\hbar}\,e^{ip^2\tau/2m\hbar}\right)\,e^{-ip^2t/2m\hbar}\right]\,e^{ipx/\hbar}\,dp 
 =  \psi(x-a,t-\tau)
\label{change_psi}
\end{equation}
so that simple shifts in coordinate and time labels are possible, 
and squeezed states often make use of similar connections.

\section{Correlated Gaussian wave packets}
\label{sec:correlated}

\subsection{Squeezed states}
\label{subsec:squeezed}

One of the simplest modifications of a standard minimum-uncertainty
Gaussian initial state which induces non-trivial initial correlations
between position and momentum is given by
\begin{equation}
\phi_{(S)}(p,0) = 
\sqrt{\frac{\alpha}{\sqrt{\pi}}}
\; e^{-\alpha^2(p-p_0)^2(1+iC)/2}
\;   e^{-ipx_0/\hbar}
\label{initial_squeezed}
\,. 
\end{equation}
(A similar version of a squeezed state, but with $\psi(x,0)$ modified,
has been discussed in Ref.~\cite{ford}.) Because the additional $C$ term
is a simple phase, the modulus of $\phi(p,t)$ is unchanged so that
the expectation values of momentum, $\langle \hat{p}\rangle_0$ and
$\langle \hat{p}^2 \rangle_0$, and the momentum-spread, are still given 
by Eqn.~(\ref{momentum_results}) as for the standard Gaussian example.
However, there is now an obvious coupling between the usual `smooth'
$\exp(-\alpha^2(p-p_0)^2/2)$ term which describes the peak momentum values
and the `oscillatory' $\exp(-ipx_0/\hbar)$ terms which dictates the
spatial location and spread, governed by the presence of the new $C$ term, 
which leads to a non-zero initial $x-p$ correlation.

The time-dependent position-space wavefunction is obtained via Fourier
transform with literally no more work than for the standard Gaussian and
one finds
\begin{equation}
\psi_{(S)}(x,t) = \frac{1}{\sqrt{\sqrt{\pi} \beta (1+i[C+t/t_0])}}
\,
e^{ip_0(x-x_0)/\hbar}
\, e^{-ip_0^2t/2m\hbar}
\,
e^{-(x-x_0-p_{0}t/m)^2/2\beta^2(1+i[C+t/t_0])}
\label{squeezed_position}
\end{equation}
giving 
\begin{equation}
|\psi_{(S)}(x,t)|^2
= \frac{1}{\sqrt{\pi}b(t)}
\, e^{-[x-\overline{x}(t)]^2/b^2(t)}
\qquad
\mbox{where}
\qquad
b(t) \equiv \beta \sqrt{1+(C+t/t_{0})^2}
\,. 
\end{equation}
Thus, the initial state in Eqn.~(\ref{initial_squeezed}) gives the same 
time-dependent Gaussian behavior as the standard case, still peaked at 
$x=\overline{x}(t)$, but with a spatial width shifted in time from 
$t \rightarrow t + Ct_0$. This can be understood from the results in 
Eqns.~(\ref{change_phi}) and (\ref{change_psi}) where  the new
$C$-dependent terms in Eqn.~(\ref{initial_squeezed}) give rise to 
effective $a$ and $\tau$ shifts given by 
\begin{equation}
a = -C\alpha^2 \hbar p_0
\qquad
\quad
\mbox{and}
\qquad
\quad
\tau = - C\alpha^2m\hbar = - Ct_0
\, . 
\end{equation}
The $\tau$ shift then affects the time-dependent width, $b(t)$, but the 
combined $a,\tau$ shifts undo each other in the argument of the Gaussian
exponential because they are highly correlated due to the form in
Eqn.~(\ref{initial_squeezed}).

The time-dependent position expectation values are then
\begin{equation}
\langle \hat{x}\rangle_t = \overline{x}(t)
\qquad
\quad
\mbox{and}
\qquad
\quad
\langle \hat{x}^2 \rangle_t = [\overline{x}(t)]^2 + \frac{[b(t)]^2}{2},
\end{equation}
so that
\begin{eqnarray}
(\Delta x_t)^2  =  \frac{[b(t)]^2}{2} 
& = &  
\frac{\beta^2}{2} \left(1 + (C+t/t_0)^2\right) \nonumber \\
& =& 
\frac{\beta^2}{2}(1+C^2) + C\beta^2\frac{t}{t_0} + \frac{\beta^2 t^2}{2t_0^2}
\nonumber \\
& = &  (\Delta x_0)^2 + At + \frac{(\Delta p_0)^2 t^2}{m^2}
\label{squeezed_spread}
\end{eqnarray}
which has a non-vanishing linear term if $C\neq 0$. The initial width
of this packet is larger than for the minimal uncertainty solution
by a factor of $\sqrt{1+C^2}$, but has the same quadratic time-dependence
since $\Delta p_0$ is the same.

One can confirm by direct calculation that $\phi_{(S)}(p,0)$ and 
$\psi_{(S)}(x,0)$ both do 
 have an initial non-vanishing
correlation leading to this form and this is also clear from plots of the
initial wave packet as shown in Fig.~2. We plot there an example with the
same model parameters as in Fig.~1, but with $C=-2$ which leads to an
anti-correlation (since $C<0$) with higher momentum components (more wiggles) 
in the `back edge' of the initial packet.  This gives an intuitive
expectation for a wave packet which
initially shrinks in time, consistent with the result in 
Eqn.~(\ref{squeezed_spread}), and with the plot shown in Fig.~2 for
$t=2t_0$. The parameters were chosen such that for this time the initial
correlation has become `undone', leading to something like the standard
Gaussian initial state, from which point it spreads in a manner which is
more familiar.  The initial correlation is achieved, however, 
at the cost of increasing the initial uncertainty principle product
by a factor of $\sqrt{1+C^2}$. 
The complete time-dependent correlation coefficient
from Eqn.~(\ref{correlation_coefficient}) is 
\begin{equation}
\rho(x,p;t) = \frac{(C+t/t_0)}{\sqrt{1+(C+t/t_0)^2}}
\end{equation}
corresponding in this case to a roughly $90\%$ initial correlation. 
The required initial correlation is also clearly evident
from the Wigner quasi-probability distribution for this case, where we
find 
\begin{equation}
P_{W}(x,p;t) = \frac{1}{\hbar \pi}
\, e^{-(p-p_0)^2 \alpha^2}
\, e^{-(x-x_0-pt/m - C(p-p_0)t_0/m)^2/\beta^2}
\,.
\end{equation}
In this case, the initial correlation for $C<0$ shown in Fig.~2 is
consistent with the desired anti-correlation, since the slope of the
elliptical contours is negative.

In a very similar manner, the expressions for the kinetic energy
density distribution from Eqn.~(\ref{define_r_function}) are simply 
shifted to
\begin{equation}
R^{(\pm)}_{(S)}(t) \equiv \frac{T^{(\pm)}_{(S)}(t)}{T^{(+)}_{(S)}(t) + T^{(-)}_{(S)}(t)}
= \frac{1}{2} \pm 
 \left(\frac{2}{\sqrt{\pi}}\right)
\left( \frac{(p_0\alpha)}{(2(p_0\alpha)^2+1)}\right) 
\frac{(C+t/t_0)}{\sqrt{1+(C+t/t_0)^2}}
\end{equation}
so that for $C<0$, there is an initial asymmetry in the front/back kinetic
energy distribution, with more `wiggles' in the trailing half of the
packet. For the $C=-2$ case in Fig.~2, the initial ($t=0$)
front/back asymmetry is $R^{(+)}/R^{(-)} = 44\%/56\%$.

We note that while a number of quantities (time-dependent spread in position,
correlation coefficient, kinetic energy distribution) are simply obtained 
by the $t \rightarrow t + Ct_0$ shift, other important metrics, such as the
autocorrelation function \cite{bassett_2}, $A(t)$, retain basically 
the same form.

One can imagine generating initial Gaussian states with non-zero 
correlations of the type in Eqn.~(\ref{initial_squeezed}), motivated 
by results obtained by the use of  modern atom trapping techniques, 
such as in Ref.~\cite{meekhof}. In a number of such experiments, 
harmonically bound ions are cooled to essentially their ground state,
after which changes in the external binding potential can generate
various {\it nonclassical motional states} such as coherent states
(by sudden shifts in the central location of the binding potential
\cite{heinzen}) and squeezed states (by changing the strength of the
harmonic binding force, i.e., the spring constant). The subsequent 
time-development of Gaussian packets in such states can then lead to
the desired correlated states, at which point the external binding
potential can be suddenly removed, with free-particle propagation
thereafter. 

As an example, the initial state in a harmonic oscillator potential
of the form $V(x) = m\omega^2x^2/2$ given by
\begin{equation}
\psi(x,0) = \frac{1}{\sqrt{\beta \sqrt{\pi}}}
\, e^{ip_0x/\hbar}\,e^{-x^2/2\beta^2}
\end{equation}
evolves in time as \cite{bassett}
\begin{equation}
\psi(x,t) = 
\exp
\left[
\frac{im\omega x^2 \cos(\omega t)}{2\hbar \sin(\omega t)}
\right]
\frac{1}{\sqrt{A(t) \sqrt{\pi}}}
\exp
\left[ 
-\frac{i m \omega \beta}{2\hbar \sin(\omega t)}
\frac{(x-x_s(t))^2}{A(t)}
\right]
\label{position_space_sho_solution}
\end{equation}
where
\begin{equation}
A(t) \equiv \beta \cos(\omega t) + i \left(\frac{\hbar}{m \omega \beta}
\right) \sin(\omega t)
\qquad
\mbox{and}
\qquad
x_s(t) \equiv \frac{p_0 \sin(\omega t)}{m \omega}
\, .
\end{equation}
The time-dependent expectation values are then
\begin{equation}
\langle x\rangle_t = x_s(t)
\, ,
\qquad
\Delta x_t = \frac{|A(t)|}{\sqrt{2}}
\, ,
\qquad
\mbox{and}
\qquad
\langle p \rangle_t = p_0\cos(\omega t)
\end{equation}
and it is then easy to show that the time-dependent correlation of this
state is given by
\begin{equation}
cov(x,p;t) = \frac{m\omega \sin(\omega t)\cos(\omega t)}{2}
\left[
\left(
\frac{\hbar}{m\omega \beta}\right)^2 - \beta^2
\right]
\,.
\end{equation}
For the special case of coherent states, where $\beta = \sqrt{\hbar/m\omega}$,
the correlations vanish identically for all times (as does the asymmetry in 
kinetic energy \cite{bassett}), while for more general solutions, removing 
the potential at times other than integral multiples of $\tau/2$ (where 
$\tau$ is the classical period) would yield an initially correlated Gaussian.

\subsection{Linear combinations of Gaussian solutions}
\label{subsec:linear_combination}

One of the simplest examples of correlated position-momentum behavior of
a system, leading to an initial shrinking of a spatial width, can be
classically modelled by two 1D non-interacting particles, with the faster 
particle placed initially behind the slower one. A quantum mechanical 
solution of the free-particle 
Schr\"{o}dinger equation involving simple Gaussian forms which mimics this quasi-classical behavior, and for which all expectation values and correlations
can be evaluated in simple closed form, consists of a linear combination
of two minimal-uncertainty Gaussian solutions of the form
\begin{equation}
\psi_{2}(x,t) = N\left[
\cos(\theta) \psi_{(G)}^{(A)}(x,t)
+ 
\sin(\theta) \psi_{(G)}^{(B)}(x,t)
\right]
\label{two_gaussians}
\end{equation}
where $A,B$ correspond to two different sets of initial position and 
momentum parameters, namely $(x_A,p_A)$ and $(x_B,p_B)$, $\theta$ describes
the relative weight of each component, and $N$ is an overall normalization;
we assume for simplicity that each component Gaussian has the same initial
width, $\beta$.
Since each $\psi_{(G)}(x,t)$ is separately normalized, the value of $N$
can be easily evaluated using standard Gaussian integrals with the result
that
\begin{equation}
N^{-2} = 
1 
+
\sin(2\theta)
\;
e^{-(x_A-x_B)^2/4\beta^2
- (p_A-p_B)^2\beta^2/4\hbar^2}
\cos[(x_B-x_A)(p_B+p_A)/2\hbar]
\end{equation}
so that if the two initial Gaussians are far apart in phase space, namely if 
\begin{equation}
\frac{(x_A-x_B)^2}{4\beta^2}
+ 
\frac{(p_A-p_B)^2\beta^2}{4\hbar^2}
>> 1
\, , 
\end{equation}
the normalization factor $N$  can be effectively set to unity, and 
all cross-terms in the evaluation of expectation values can also 
be neglected. 

In this limit, the various initial expectation values required for the
evaluation of the time-dependent spread in Eqn.~(\ref{general_case}) are 
given by 
\begin{eqnarray}
\langle \hat{x} \rangle_0 & = & \cos^2(\theta) x_A + \sin^2(\theta) x_B
\\
\langle \hat{x}^2 \rangle_0 & = &
\cos^2(\theta) \left(x_A^2 + \frac{\beta^2}{2}\right)
+ 
\sin^2(\theta) \left(x_B^2 + \frac{\beta^2}{2}\right)
- \left[\cos^2(\theta) x_A + \sin^2(\theta) x_B\right]^2
\end{eqnarray}
so that
\begin{equation}
(\Delta x_0)^2 = 
[\sin(2\theta)]^2 \left(\frac{x_A-x_B}{2}\right)^2
+ \frac{\beta^2}{2}
\end{equation}
with a similar result for the momentum-spread, namely
\begin{equation}
(\Delta p_0)^2 = 
[\sin(2\theta)]^2 \left(\frac{p_A-p_B}{2}\right)^2
+ \frac{\hbar^2}{2\beta^2}\,.
\end{equation}
The necessary initial correlation is given by
\begin{equation}
\langle \hat{x}\hat{p} + \hat{p}\hat{x} \rangle_0 - 
2\langle \hat{x}\rangle_0 \langle \hat{p} \rangle_0
= 
2 [\sin(2\theta)]^2 \left[\frac{(x_A-x_B)(p_A-p_B)}{4}\right]
\end{equation}
so that the time-dependent spread in position is given by
\begin{equation}
(\Delta x_t)^2 =
[\sin(2\theta)]^2 
\left[
\left(\frac{x_A-x_B}{2}\right)
+ \left(\frac{p_A-p_B}{2}\right)\frac{t}{m}
\right]^2
+
\frac{\beta^2}{2}
+
\frac{\hbar^2 t^2}{2m^2\beta^2}
\,. 
\end{equation}
In the limit we're considering, namely when $|x_A-x_B| >> \beta$
and/or $|p_A-p_B| >> \hbar/\beta$, the time-dependent width can be dominated
by the quasi-classical value dictated by two well-separated `lumps' of
probability, and if $(x_A-x_B)$ and $(p_A-p_B)$ have opposite signs, then this
large position spread can initially decrease in time because of the
initial correlations. This example, while not as `quantum mechanical' as
that in Sec.~\ref{subsec:squeezed}, does clearly and simply  exhibit the 
position-momentum correlations necessary for the presence of the $A$ term 
in Eqn.~(\ref{squeezed_spread}), with the `fast one in the 
back, and the slow one in the front'.

One can imagine producing linear combinations of isolated, but highly 
correlated, Gaussian wave packets at very different points in phase space, 
by invoking the dynamical time-evolution of bound state wave packets which 
leads to the phenomenon of wave packet revivals, especially
fractional revivals \cite{revivals}. For the idealized case of the
infinite square well potential \cite{aronstein}, 
at $t=T_{rev}/4$ (where $T_{rev}$ is
the full revival time), an initially localized wave packet is 'split'
into two smaller copies of the original packet, located at opposite
ends of phase space \cite{belloni}, of the form in Eqn.~(\ref{two_gaussians}).
If, in this model system, the infinite wall boundaries are suddenly
removed at such a point in time, 
we then have the case considered in this section.

\section{Conclusion and discussion}
\label{sec:conclusion}

The study of the time-dependence of the spatial width of wave packets
in model systems can produce many interesting results, a number of which
are quasi-classical in origin, while some are explicitly quantum mechanical.
Time-dependent wave packet solutions of the Schr\"{o}dinger equation for
the harmonic oscillator are easily shown to exhibit intricate correlated
expansion/contraction of widths in position- and momentum-space 
\cite{saxon} and modern experiments \cite{meekhof}, \cite{heinzen} 
can probe a wide variety of such states. 
Even the behavior of otherwise free Gaussian wavepackets 
interacting with (or `bouncing from')  an infinite wall 
\cite{doncheski_1}, \cite{dodonov}, \cite{doncheski_2}
can lead to wave packets which temporarily shrink in size.

While the fact that free-particle wavepackets can also exhibit 
initial shrinking of their spatial width is well-known in the
physics pedagogical literature, it is perhaps not appreciated enough 
in the context of introductory quantum mechanics courses because of the 
seeming lack of simple, mathematically tractable, and intuitively
visualizable examples, and we have provided two such simple cases here. 
We have also emphasized the usefulness of several tools for the detailed 
analysis of the structure of quantum states as they evolve, namely the direct
visualization of the real/imaginary part of the spatial wavefunction, the
time-dependent spatial distribution of the kinetic energy (how the
`wiggliness' changes in time), and the Wigner quasi-probability
distribution all of which provide insight into
the correlated $x-p$ structure of quantum states.

\appendix
\section{Time-dependent $\Delta x_t$ for the case of uniform acceleration}
\label{sec:appendix}

Many of the results discussed here  for the time-dependent widths of
localized quantum wavepackets for the free-particle case can be carried
over to the situation of a particle undergoing uniform acceleration, 
governed by the Hamiltonian $\hat{H} = \hat{p}^2/2m -Fx$, corresponding
to a constant force, $+F$, to the right. General expressions for the 
time-dependent values of expectation values of powers of both $x$ and $p$ 
can be obtained using Eqn.~(\ref{time-development}) (following the method of 
Styer \cite{styer} for example) and one obtains the results
\begin{equation}
\langle \hat{p} \rangle_t  =  Ft + \langle p \rangle_0 
\qquad
\quad
\mbox{and}
\quad
\qquad
\langle \hat{p}^2 \rangle_t  =  F^2t^2 + 2F\langle \hat{p}\rangle_0 t +
\langle \hat{p}^2 \rangle_0
\label{acc_p_1}
\end{equation}
which imply that 
$(\Delta p_t)^2 = (\Delta p_0)^2$. 
For position we have the corresponding results
\begin{eqnarray}
\langle \hat{x} \rangle_t 
& = & \frac{Ft^2}{2m} + \frac{\langle \hat{p} \rangle_0t}{m}
+ \langle \hat{x} \rangle_0 
\label{acc_x_1} \\
\langle \hat{x}^2 \rangle_t
& = &
  \frac{F^2t^4}{4m^2} 
+ \frac{F\langle \hat{p}\rangle_0 t^3}{m^2}
+ \frac{\langle \hat{p}^2\rangle_0 t^2}{m^2}
+ \frac{F\langle \hat{x}\rangle_0 t^2}{m}
+ \frac{\langle \hat{x}\hat{p} + \hat{p}\hat{x} \rangle_0t}{m}
+ \langle \hat{x}^2 \rangle _0 
\label{acc_x_2}
\end{eqnarray}
which actually combine to give the same expression for $(\Delta x_t)^2$ as in 
Eqn.~(\ref{general_case}). Thus, the model systems we have discussed
here can also be used as examples of correlated wave packets in the
related accelerating particle case.

The expressions above can also be derived using an approach similar to that
followed in Sec.~\ref{sec:momentum_space}, namely by using the most
general form for the time-dependent momentum-space wavefunction, $\phi(p,t)$.
In that case, the time-dependent Schr\"{o}dinger equation in $p$-space is 
written in the form
\begin{equation}
\frac{p^2}{2m} \phi(p,t)
+ F \left[i\hbar \frac{\partial}{\partial p}\right]
\phi(p,t) 
= i \hbar \frac{\partial}{\partial t} \phi(p,t)
\end{equation}
which has a general solution
\begin{equation}
\phi(p,t) = \phi_{0}(p-Ft)\, e^{i[(p-Ft)^3-p^3]/6mF\hbar}
\end{equation}
where $\phi_{0}(p) = \phi(p,0)$ is still the initial wavefunction.
The $t$-dependent expectation values in Eqn.~(\ref{acc_p_1}), 
(\ref{acc_x_1}) and (\ref{acc_x_2}) can then be obtained as in 
Sec.~\ref{sec:momentum_space} in terms of the $t=0$ results, just as
for Eqns.~(\ref{p_1}), (\ref{p_2}), (\ref{x_1}), and
(\ref{x_2}).

 \newpage

\begin{flushleft}
{\Large {\bf 
Figure Captions}}
\end{flushleft}
\vskip 0.5cm
 
\begin{itemize}
\item[Fig.\thinspace 1.]  The time-development of a standard Gaussian
wavepacket, $\psi_{(G)}(x,t)$, described by 
Eqn.~(\ref{free_particle_position_solution}),  with the 
modulus (solid) and real part (dashed) shown for $t=0$ and $t=2t_0$
on the top plot. Contour plots of the Wigner function, from 
Eqn.~(\ref{explicit_wigner_function}), for the same two times are 
shown at the bottom,
corresponding to contours which are $70\%$, $30\%$, and $10\%$ of the
peak or central value. The model parameters used in this plot are 
$\hbar = m = 1$ and $\beta = 2$, which give $t_0 \equiv m\beta^2/\hbar= 4$, 
along with $p_0 = 4$ and $x_0 = -4$ for the initial packet.
\item[Fig.\thinspace 2.]  Same as Fig.~1, but for the correlated
squeezed state in Eqn.~(\ref{initial_squeezed}) or (\ref{squeezed_position}),
with the model parameters used in Fig.~1. For this case, we use $C=-2$, so 
that for $t=2t_0$ the initial correlations are `undone'.
\end{itemize}

\newpage

\begin{figure}[hbt]
\begin{center}
\epsfig{file=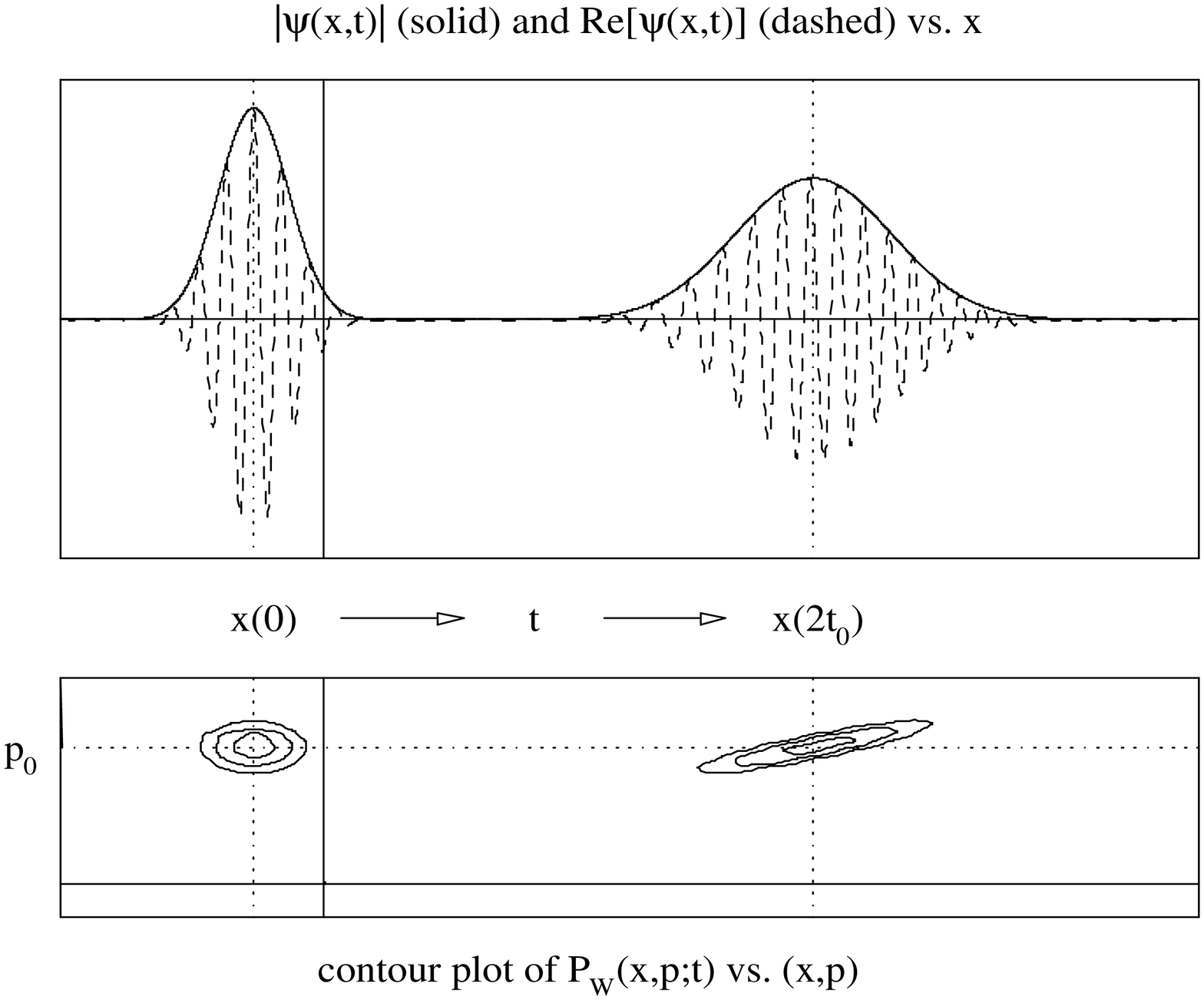,width=0.8\linewidth,angle=270}
\caption{}
\end{center}
\end{figure}

\newpage

\begin{figure}[hbt]
\begin{center}
\epsfig{file=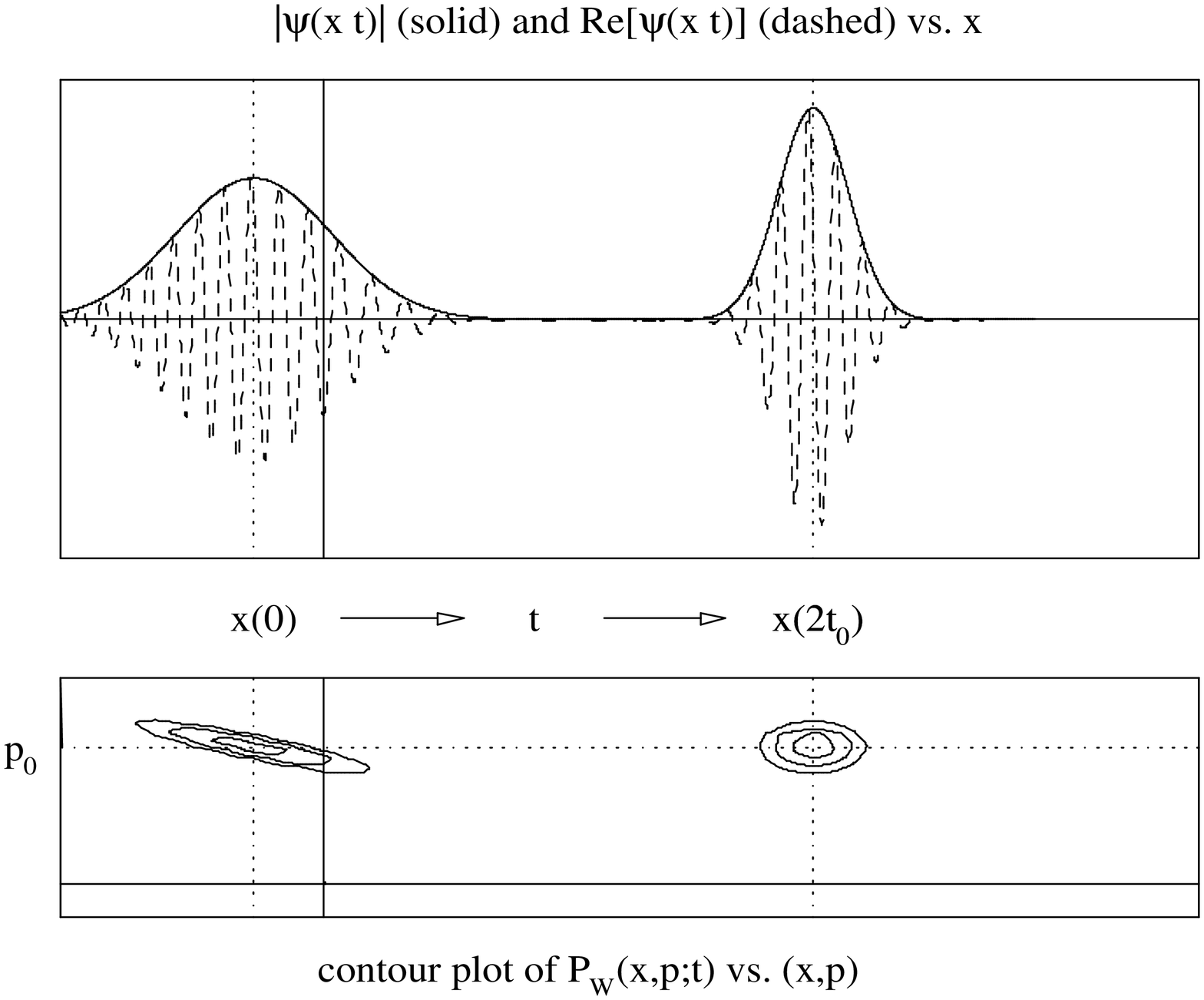,width=0.8\linewidth,angle=270}
\caption{}
\end{center}
\end{figure}

\end{document}